# Anomalous Nernst Effect and Its Implications for Time-Reversal Symmetry Breaking in Kagome Metal ScV$_6$Sn$_6$

Yazhou Li[1,#], Saizheng Cao[2,#], Jiaxing Liao[1], Jiajun Ma[1], Yuwei Zhang[1], Tao Li[1], Jialu Wang[1], Chenchao Xu[1], Jianhui Dai[1], Chao Cao[2], Yu Song[&,2], Peijie Sun[3], Yuke Li[*,1]

1. School of Physics and Hangzhou Key Laboratory of Quantum Matters, Hangzhou Normal University, 311121, China
2. Center for Correlated Matter and School of Physics, Zhejiang University, Hangzhou, 310058, China.
3. Beijing National Laboratory for Condensed Matter Physics, Institute of Physics, Chinese Academy of Sciences, Beijing, China

## Abstract

The nonmagnetic kagome metal ScV$_6$Sn$_6$ displays an unconventional charge order (CO) accompanied by signatures of an anomalous Hall effect, hidden magnetism, and multiple lattice instabilities. In this study, we report the observation of unconventional anomalous thermoelectric properties. Notably, unexpected anomalous transverse Nernst signals reach a peak value of ~4 µV/K near the $T_{CDW}$ ~92 K in ScV$_6$Sn$_6$, and these signals persist in the charge-ordered state as the temperature decreases to 10 K. Furthermore, both thermopower and thermal conductivity exhibit significant changes under magnetic fields, even in the nonmagnetic ground state. These observations strongly suggest the emergence of time-reversal symmetry breaking in ScV$_6$Sn$_6$, as supported by muon spin relaxation (µSR) measurements. While hidden magnetism represents the most plausible origin, alternative mechanisms involving orbital currents and chiral charge order remain possible.

## Introduction

Kagome lattice with its unique geometry, giving a unique electronic structure with Dirac points, flat bands, and van Hove singularities (VHS), has made it a prominent platform for exploring the interplay between topology, charge, and magnetism[1-3]. Recently, a large number of kagome-based materials have been identified that exhibit many-body quantum phenomena, including superconductivity[4], charge-density-wave (CDW)[5, 6], chiral current[7], and anomalous Hall/Nernst effect[8-10]. Among those materials, the AV$_3$Sb$_5$ (A =K, Rb, Cs) family[4, 11] has attracted significant attention due to its potential to harbor various competing orders, including superconductivity, CDW, and topological states. Another example is the antiferromagnetic kagome compound FeGe[5, 12-14], which displays enhanced iron moments associated with short-range CDW correlations within a collinear antiferromagnetic (AFM) order. Notably, disorder can convert short-range CDW into long-range order[13, 15, 16]. Therefore, the various instabilities associated with charge and spin, as well as their interplay, play a crucial role in the complexity and fertility of physical properties in kagome materials.

Very recently, an unconventional CDW phase with a wave vector (1/3,1/3,1/3) around 92 K was found in the nonmagnetic Kagome metal $ScV_6Sn_6$ [6, 17-21], belonging to the kagome family of the hexagonal $HfFe_6Ge_6$-type compounds which have been extensively studied for their magnetism and topological properties, but typically without exhibiting CDW order[22]. Unlike the in-plane vanadium displacements of the kagome sublattice in $AV_3Sb_5$[4], the CDW transition in $ScV_6Sn_6$ is tied to the out-of-plane displacement of Sc and Sn atoms[18]. Optical spectroscopy[23] and angle-resolved photoemission spectroscopy (ARPES) [24] measurements indicate that Fermi-surface nesting and a significant charge gap related to the CDW are absent in $ScV_6Sn_6$. Muon spin resonance (μSR) spectroscopy[25] suggests the presence of hidden magnetism, leading to time-reversal symmetry breaking in the CDW state, which supports the observed anomalous Hall effects[26, 27]. The lack of an anomalous Hall effect-like feature in chromium-substituted samples[28] further reinforces the idea of hidden magnetism being associated with the CDW phase. Inelastic X-ray scattering and X-ray diffraction experiments[19, 21, 29], along with first-principles calculations[17, 30], have revealed the existence of another dynamic short-range CDW phase with a wave vector of (1/3, 1/3, 1/2) above 92 K[17-21]. This phase appears to compete with the long-range CDW, indicating that strong electron-phonon coupling[31] plays a crucial role in the formation of the CDW state, as well as the magnitude of transport coefficients, including anomalous Hall and Nernst effect. Despite these insights, the origins of unconventional transport properties in $ScV_6Sn_6$, such as the anomalous Hall effect (AHE), remain debated. Notably, the anomalous thermoelectric effects—including the Seebeck and Nernst effects—and their intrinsic relationship with charge orders have yet to be systematically explored. Given that charge order often intertwines with electronic symmetry breaking, the potential role of time-reversal symmetry breaking (TRSB) in these phenomena presents a critical open question. TRSB in $ScV_6Sn_6$, whether driven by emergent/hidden magnetic order and chiral charge order, directly influences anomalous transverse transport. Resolving this connection among TRSB, charge order, and thermoelectric effects is crucial to unraveling the microscopic mechanisms behind $ScV_6Sn_6$'s exotic electronic behavior.

Here, we present a detailed study of anomalous thermal transport properties in $ScV_6Sn_6$. Below $T_{CDW}$ within the nonmagnetic ground state in $ScV_6Sn_6$, the resistivity, thermopower, and thermal conductivity exhibit unusually pronounced sensitivities to the applied magnetic fields, indicating a signature of hidden magnetism. Importantly, a step-like Nernst signal observed confirms an anomalous Nernst component, reaching a maximum value of approximately 4 μV/K near $T_{CDW}$. The ANE can be ascribed to chiral current effects or hidden magnetism breaking time-reversal symmetry. Additionally, the anomalous Nernst signals survive above $T_{CDW}$, possibly associated with charge-order fluctuations. Our findings reinforce the observation of anomalous thermoelectric effect and its implication for time-reversal symmetry broken in the non-magnetic $ScV_6Sn_6$ and highlight the intricate correlation between charge order, topological bands, and hidden magnetism.

## Experimental methods

High-quality single crystals of $ScV_6Sn_6$ were grown using the self-flux method, as previously reported[21]. The majority of as-grown crystals exhibit plate-like morphology with the c-axis perpendicular to the largest surfaces(ab-plane), displaying typical dimensions of around 1×1×0.5 $mm^3$, while a few crystals with the largest size of 0.5×0.5×1.1 $mm^3$ exhibit an elongated grown c-axis [more details are found in SI]. This pronounced morphological anisotropy facilitates investigation of anisotropic transport properties through current/heat flow configurations along both in-plane (ab) and out-of-plane (c-axis) directions. The measuring crystals are selected from the same batch. They are polished and cut into a bar shape with c-direction [see more details in SI]. The (magneto)resistivity and Hall resistivity measurements were simultaneously performed using the six-terminal method in a commercial 14 T-cryogenic refrigerator. The magnetization was measured using a commercial PPMS-VSM. The thermopower and Nernst effect were simultaneously performed with a one-heater–two-thermometer technique in a 14 T-cryogenic refrigerator with a high-vacuum environment.

In transverse transport measurements, such as Hall and Nernst effect experiments, slight misalignment of the voltage contacts can introduce an additional contribution related to the longitudinal signals (i.e., resistivity and thermopower $S_{zz}$). Since transverse transport signals are generally odd with respect to the magnetic field, the transverse Hall and Nernst signals were antisymmetrized to eliminate the effects of electrode misalignment.

## Results and discussion

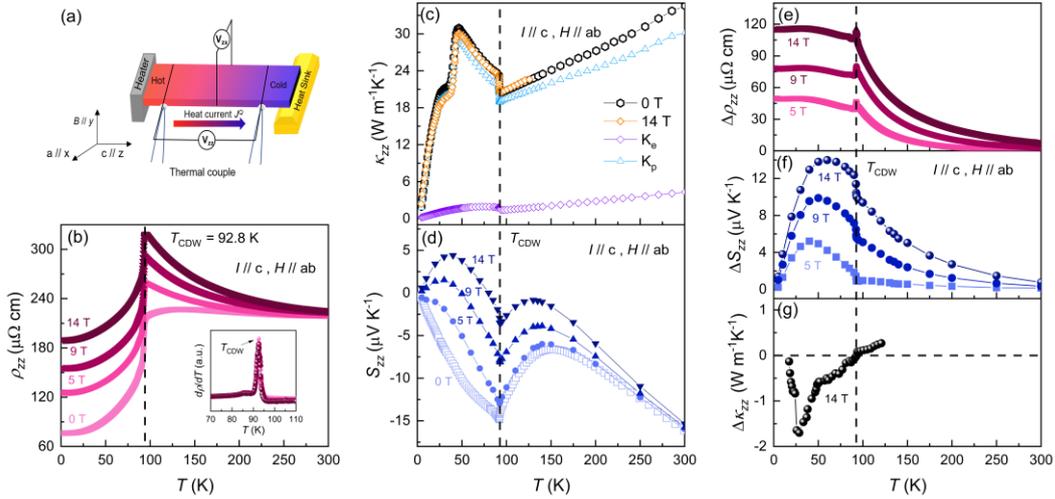

Figure 1 (a) The sketch of thermal/electrical transport measurement setup. (b) The resistivity $\rho_{zz}$ as a function of temperature under several different magnetic fields up to 14 T. The inset shows the differentiation of resistivity around $T_{CDW}$. (c) Temperature dependence of longitudinal thermal conductivity and Seebeck coefficient (d). (e) Magneto-resistivity ($\Delta\rho_{zz} = \rho_{zz}(B) - \rho_{zz}(0)$), (f) magneto-thermopower ($\Delta S_{zz} = S_{zz}(B) - S_{zz}(0)$), and (g) magneto-thermal-conductivity ($\Delta\kappa_{zz} = \kappa_{zz}(B) - \kappa_{zz}(0)$) at several magnetic fields.

The zero-field resistivity $\rho_{zz}$ in Figure 1b is almost unchanged above $T_{CDW}$, but undergoes a significant drop at 92 K owing to reduced electron-phonon scattering resulting from the opening of a gap in the Fermi surface, marking the CDW transition. Interestingly, $\rho_{zz}$ shows a saturated resistivity plateau at lower temperatures in Figure 1b, which is generally ascribed to the surface state observed in previous topological semimetals, like TaSb$_2$[32]. Studies of quantum oscillations[26,

33, 34] also reveal the presence of a nontrivial Dirac band in ScV$_6$Sn$_6$ [See more details in SI]. The plateau in $\rho_{zz}$ is robust against magnetic fields, persisting up to 14T. The application of magnetic fields does not shift $T_{CDW}$ (see the inset of Figure 1b), but substantially enhances the resistivity values across the entire temperature region. Consequently, we observe a field-induced semiconductor-like behavior in $\rho_{zz}$ for $B > 5$ T. We extract the magneto-resistivity (MR) components for magnetic fields of 5, 9, and 14T, as illustrated in Figure 1e. It is clearly seen that the change of resistivity, $\Delta\rho_{zz}$, increases rapidly with decreasing temperature towards $T_{CDW}$, reaching a maximum value at $T_{CDW}$ and subsequently showing a clear resistivity-plateau at low temperatures. This feature indicates that the dominant scattering mechanisms above and below $T_{CDW}$ may differ. The unusual MR behaviors observed in ScV$_6$Sn$_6$ contrast sharply with those in other kagome metals, such as the paramagnetic CsV$_3$Sb$_5$ and antiferromagnetic (AFM) FeGe with CDW[5, 13, 15], as well as the isostructural LuV$_6$Sn$_6$, which lacks a CDW[27]. This disparity indicates that the distinctive CDW mechanism in ScV$_6$Sn$_6$, possibly linked to other orders like hidden magnetism[25], may play a key role in determining its transport properties.

The thermal-conductivity $\kappa_{zz}$, illustrated in Figure 1c, demonstrates the difference that $\kappa_{zz}$ follows a linear decrease with cooling temperature above $T_{CDW}$, then shows a sharp increase related to structural distortion along the c-axis due to the CDW transition. This behavior is different from the ignorable CDW anomaly in thermal conductivity observed in CsV$_3$Sb$_5$[9]. Applying a magnetic field of 14 T significantly suppresses a shoulder anomaly in zero-field $\kappa_{zz}$ at 25 K, shifting it to lower temperatures. This field-dependent anomaly is hard to explain by the phonons and electrons scattering. We thus can separate these contributions from the total thermal conductivity in the nonmagnetic ScV$_6$Sn$_6$ system, which is represented as $\kappa_{zz} = \kappa_e + \kappa_{ph}$ where $\kappa_e$ is the electron thermal conductivity and $\kappa_{ph}$ represents the phonon contribution. The $\kappa_e$ can be calculated using the Wiedemann-Franz law, given by $\kappa_e = L_0 \sigma_{zz} T$, where $L_0 = 2.44 \times 10^{-8} W\Omega/K^2$ is the Lorentz number and $\sigma_{zz}$ is the longitudinal conductivity along c-direction, as shown in Figure 1c. The obtained $\kappa_e$, which does not exhibit the signature of anomaly near 25 K, is nearly an order of magnitude smaller than the total $\kappa_{zz}$, yielding a comparable large $\kappa_{ph}$. This implies that phonons dominate thermal transport [31], suggesting that the CDW primarily originates from structural factors rather than electronic instabilities [19], unlike the case in CsV$_3$Sb$_5$ [7]. Note the Lorentz number exhibits strong temperature dependence and deviates significantly from $L_0$ below $T_{CDW}$ in AV$_3$Sb$_5$ (A= K, Cs)[35]. A similar behavior is also observed in ScV$_6$Sn$_6$, but the derived electron thermal conductivity remains smaller than $\kappa_{ph}$ (More detailed analysis in SI). The linear temperature dependence of thermal conductivity above $T_{CDW}$ can be attributed to charge fluctuations, analogous to the behavior observed in $AV_3Sb_5$ (A= K, Cs)[35]. Analyzing the difference $\Delta\kappa_{zz} = \kappa_{zz}$ (14T) $-\kappa_{zz}$ (0T) in Figure 1g reveals a slight decrease below $T_{CDW}$, dipping to a minimum of 1-2 W/m K at 25 K. Given the robustness of $\kappa_{ph}$ against magnetic fields and the small electron components $\kappa_e$, we conjecture that the finite $\Delta\kappa_{zz}$ may arise from contributions related to a hidden magnetic order below $T_{CDW}$, as detected in μSR experiments[25]. Other factors, including charge fluctuations and electron-phonon coupling, may also be involved.

Thermopower $S_{zz}$ as a function of temperature in Figure 1d is negative within the measured regime, indicating the majority of electron-type carriers. Its absolute value, $|S_{zz}(T)|$ shows a linear temperature dependence above 150 K, followed by a broad hump at 150 K, and then a deep valley at $T_{CDW}$. As the temperature decreases further, $|S_{zz}|$ turns to decrease linearly to approximate zero at the lowest temperature. Applying a magnetic field substantially suppresses the thermopower, especially below $T_{CDW}$, while also introducing a broad anomaly with a maximum value of around 25 K, similar to the observation in $\kappa_{zz}$ (See Figures 1c and 1g). Increasing magnetic fields from 5 to 14 T enhances this anomaly and causes the sign of $S_{zz}$ to change from negative to positive below $T_{CDW}$. Interestingly, compared with the robustness of $T_{CDW}$ against magnetic fields, the hump observed above $T_{CDW}$ shifts to low temperatures. In contrast, the field-induced anomaly below $T_{CDW}$ is pushed towards higher temperatures with increasing magnetic fields, suggesting a possible contribution linked to the hidden magnetic order.

The unconventional behaviors observed in thermopower across the CDW transition can be further illustrated by the magneto-thermopower $\Delta S_{zz} = S_{zz}(14T) - S_{zz}(0T)$ in Figure 1f. Above $T_{CDW}$, the aforementioned hump in $S_{zz}$ vanishes in the $\Delta S_{zz}$ curves. Instead, $\Delta S_{zz}$ shows a "semiconductor-like" behavior as a function of temperature, resembling the features observed in $\Delta\rho_{zz}$. Below $T_{CDW}$, $\Delta S_{zz}$ obeys a linear decrease with temperature following the anomaly, while in the temperature regime, $\Delta S_{zz}$ remains constant. The unconventional behaviors in $\Delta S_{zz}$ after and before $T_{CDW}$ imply that the enhanced anomaly under magnetic fields may arise from the contributions of hidden magnetic order in the charge ordering states.

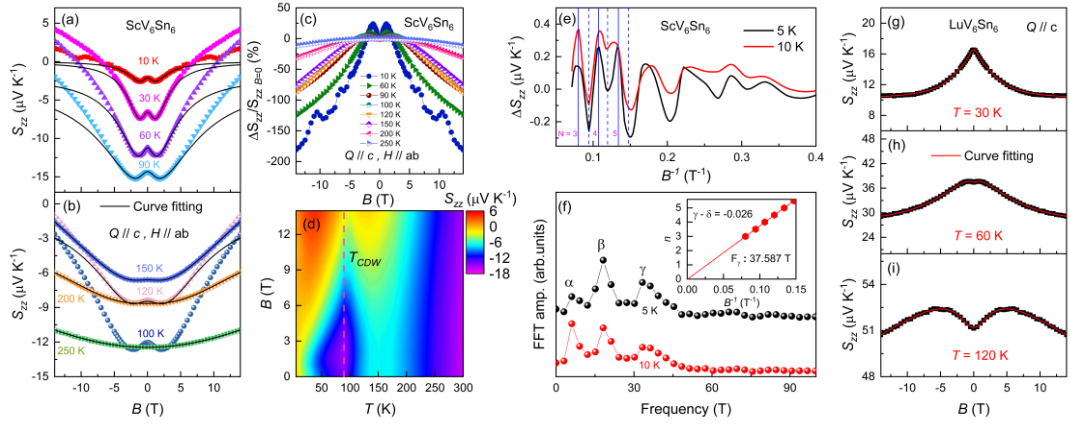

Figure 2 (a)-(b) Thermopower $S_{zz}$ as a function of magnetic fields at different temperatures. (c) Magneto-thermopower vs. magnetic fields. (d) The contour plot of thermopower as a function of magnetic field and temperature. (e) Oscillatory components of $S_{zz}$ at different temperatures. (f) FFT spectrum of the Seebeck with the peaks labeled. The inset shows the Landau index $n$ vs $B^{-1}$. (g)-(i) Thermopower $S_{zz}$ in the isostructural LuV$_6$Sn$_6$ as a function of magnetic fields at 30 K, 60 K, and 120 K, respectively. The red line represents the fitting curves using the two-band model.

Thermopower $S_{zz}$, as shown in Figure 2, has a mustache-shaped profile at low fields, with a minimum value occurring around the crossover field $B_{cr}$. This feature may be explained by the multi-band nature of the system. Below $T_{CDW}$, the profile becomes more prominent with temperature increasing to $T_{CDW}$, while $B_{cr}$ slowly shifts to higher fields, similar to the trend observed in MR data (See Figure S2c in SI). After the $B_{cr}$, $S_{zz}$ starts to increase and exhibits a linear dependence on the magnetic field at high fields without saturation. Note that a field-induced sign-reversal in $S_{zz}$ can be observed, with the sign-reversal field increasing substantially at higher temperatures. As $T > T_{CDW}$, the mustache-shaped profile in $S_{zz}$ gradually broadens and eventually disappears over 150 K. In this high-temperature regime, $S_{zz}$ follows a parabolic dependence on the magnetic field instead of a linear relationship. Correspondingly, magneto-thermopower $MS = [S_{zz}(B) - S_{zz}(0)] / S_{zz}(0)$ reaches nearly 200 % at 10 K, significantly surpassing the MR values of 150 %. This large magneto-thermopower is relatively uncommon in the known kagome topological metals, highlighting a distinctive feature of this system.

Theoretically, the low field $S_{zz}$ signals can be explained using the conventional semiclassical expression[36].

$$S_{zz}(B) = S_1 \frac{1}{1+(\mu_1 B)^2} + S_2 \frac{1}{1+(\mu_2 B)^2} + S_\infty \frac{(\mu' B)^2}{1+(\mu' B)^2}$$

Here, $S_1(S_2)$ and $\mu_1(\mu_2)$ are the zero-field Seebeck coefficients and mobility of the two-band carrier, respectively, and $S_\infty$ is the limiting value of the thermopower when $B \gg 1/\mu$. For $ScV_6Sn_6$, the low-field $S_{zz}$ can be fitted well using the two-band model. However, at higher fields, the fitting results tend to saturate, seriously deviating from the expected linear field-dependent of $S_{zz}$ observed below $T_{CDW}$, as shown in Figure 2a. In contrast, the isostructural $LuV_6Sn_6$ without a CDW displays thermopower saturation at high magnetic fields below $T_{CDW}$, as shown in Figure 2, consistent with conventional multiband transport behavior that can be well described by a two-band model. This contrasting behavior between these isostructural compounds suggests that nontrivial topological effects, including Berry curvature, are involved in $ScV_6Sn_6$'s CDW state. Above $T_{CDW}$, the $S_{zz}$ data perfectly aligns with the fitted curves from the multiband model[37-39], indicating the distinct thermopower origins after and before the $T_{CDW}$ transition.

The deviation of $S_{zz}$ from the two-band model below $T_{CDW}$ may result from nontrivial topological bands that appear inside the CDW phase, which is supported by quantum oscillation measurements. Figure 2e shows the amplitude of oscillations $\Delta S_{zz}$ as a function of $1/B$. The periodic oscillations in $1/B$ reflect the quantization of energy levels. After performing a fast Fourier transform on the data shown in Figure 3f, three distinct frequencies were identified: $F_\alpha = 6$ T, $F_\beta = 18$ T, and $F_\gamma = 33$ T, which closely match those frequencies obtained from the Shubnikov-de Haas (SdH) oscillations[26, 33][ More details in SI], suggesting similar underlying Fermi surface topology. Based on the Lifshitz-Kosevich equation[40], the Berry phase associated with the nontrivial topology of the band can be estimated using a Landau fan diagram, as shown in the inset of Figure 2f. The Landau levels (LL) are indexed, where integer $n$th is assigned to the maxima and the LL half-integer ($n$ + 1/2) th is assigned a minimum of $\Delta S_{zz}$ (Figure 2e). We focused on a magnetic field window between 5 and 14 T to estimate the Berry phase of $F_\gamma$. The Landau index $n(=\frac{F}{B}+\gamma-\delta)$ as a function of $1/B$ show a linear behavior in Figure 2f, where the slope denotes the oscillatory frequency and the intercept offers the information of Berry phase. Accordingly, the obtained phase factor $\gamma - \delta$ was estimated to be -0.026, indicating the presence of a nontrivial Berry phase, which is part of the apple-shaped FS related to the Dirac bands[26] in the CDW phase of $ScV_6Sn_6$.

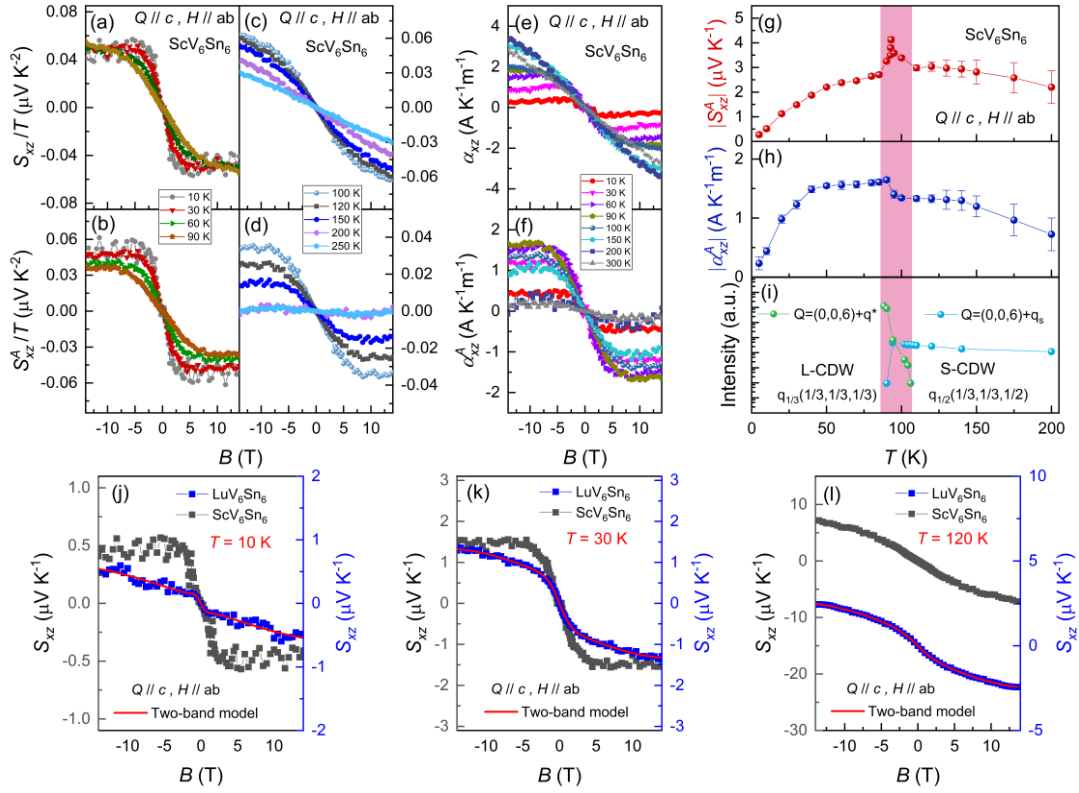

Figure 3 (a), (c) The Nernst signals $S_{xz}/T$ as a function of magnetic fields at selected temperatures across $T_{CDW}$ in ScV$_6$Sn$_6$. (b), (d) Anomalous Nernst signals. (e)-(f) Transverse thermal electrical conductivity and its anomalous components. Temperature dependence of anomalous Nernst signals (g) and thermal-electric conductivity $\alpha_{xz}^A$ (h). (i) The integrated intensities for q*-CDW representing the long-range CDW order (L-CDW,) and q$_s$-CDW denoting the short-range CDW (S-CDW,) using inelastic X-ray scattering. The data are obtained from the Ref.[21]. (j)-(l) A comparison of the Nernst signal vs. magnetic fields between ScV$_6$Sn$_6$ and LuV$_6$Sn$_6$ at selected temperatures. The red lines illustrate the fitting results from the two-band model.

The Nernst effect is more sensitive to anomalous contributions because of the derivative of the conductivities as given by the Mott relation. The magnetic field dependence of Nernst signals $S_{xz}/T$ for ScV$_6$Sn$_6$ at selected temperatures is shown in Figure 3a-3b. In the semiclassical theory, the conventional Nernst signal under magnetic fields can be expressed as $S_{xz} = S_0 \mu B / [1+(\mu B)^2]$, where $\mu$ denotes the carrier mobility. It predicts a sharp Drude-like peak at $B = 1/\mu$, followed by a decrease towards zero at higher fields. In contrast, the ANE signal rises steeply to a maximum value and then holds a plateau at high fields. In the case of ScV$_6$Sn$_6$, the Nernst signals evolving with fields exhibit a clear step-like profile at low temperatures (Figure 4a), indicating an obvious anomalous Nernst effect, possibly associated with the hidden magnetic orders. A similar thermoelectric phenomenon has been reported in CsV$_3$Sb$_5$, which is ascribed to the CDW-modulated nontrivial band structure[40]. Unlike ScV$_6$Sn$_6$, the isostructural counterpart LuV$_6$Sn$_6$ exhibits a characteristic S-shaped Nernst signal at low fields below $T_{CDW}$, as illustrated in Figure 3. This feature serves as a clear signature of conventional multiband charge transport behavior[37-39]. The stark difference between these isostructural compounds provides compelling evidence that the ANE in ScV$_6$Sn$_6$ originates from hidden magnetism or chiral charge order, inducing TRSB.

As the temperature increases, the anomalous component reaches a maximum value of ~ 4 μV/K at $T_{CDW}$ and then gradually decreases at higher temperatures (Figure 3b). Over 150 K, the $S_{xz}(B)$ exhibits a linear behavior with a negative slope, suggestive of a multiband effect. By subtracting the linear term (representing the normal Nernst effect), we can extract the anomalous Nernst components, as shown in Figure 3c-3d [See more details in SI]. The ANE $S_{xz}^A / T$ shows a saturated plateau at high fields below $T_{CDW}$, transitioning to an S-shape above 150 K. The obtained $S_{xz}^A$ as a function of temperature in Figure 3g shows a distinct peak at $T_{CDW}$, gradually decreasing to near zero as temperature approaches either zero or room temperature. The observed anomalous Nernst effect strongly suggests time-reversal symmetry breaking mediated by hidden magnetism in the unconventional CDW phase, though the potential mechanism involving chiral charge order or orbital currents requires further investigation to be ruled out. Our findings highlight the intricate interplay between the thermoelectric response and its distinct electronic structures[41].

The off-diagonal thermal-electric conductivity $\alpha_{xz}$ is closely associated with both $S_{xz}$ and $\sigma_{xz}$ via $\sigma_{zz}$ and $S_{zz}$ through the relationship[41]: $\alpha_{xz} = \sigma_{xz} S_{zz} + S_{xz} \sigma_{zz}$, as shown in Figure 3e [See more $\alpha_{xz}$ data in SI]. The evolution of $\alpha_{xz}$ with magnetic fields exhibits a clear step-like behavior at low temperatures, indicative of the anomalous components. To isolate the anomalous components $\alpha_{xz}^A$, we subtract the linear term from the total $\alpha_{xz}$, as shown in Figure 3f. Temperature dependence of $\alpha_{xz}^A$ in Figure 3h shows a pronounced peak around $T_{CDW}$. The maximum $\alpha_{xz}^A$ reaches around 1.6 A/K·m, which is significantly larger than the values typically found in conventional FM metals[42], and is comparable to those reported for various topological magnets, like CoMnGa$_2$[43], Co$_3$Sn$_2$S$_2$[8], and YMn$_6$Sn$_6$[44]. The enhanced $\alpha_{xz}^A$ reflects the intricate interplay between the ANE and the unconventional CDW order, suggesting that the topological characteristics of the material play a substantial role in its thermoelectric properties.

The observed anomalies in $S_{xz}^A$ and $\alpha_{xz}^A$ across the $T_{CDW}$ align well with the competing CDW instabilities around $T_{CDW}$ that have been identified in inelastic X-ray scattering experiments[21]. Below $T_{CDW}$, the (1/3,1/3,1/3) long-range charge density wave (L-CDW) order dominates, while the dynamic (1/3,1/3,1/2) short-range charge density wave (S-CDW) prevails above $T_{CDW}$ in ScV$_6$Sn$_6$. Within the CDW transition region between 90 K and 120 K, both the L-CDW and S-CDW coexist, leading to complex interplay. As temperature increases to near 150 K, the ANE remains discernible, which may be ascribed to the dynamic fluctuation of the CDW[45, 46]. These fluctuations are degenerate and highly mobile, meaning that local TRSB could still emerge owing to the chiral configuration of S-CDW. Consequently, the ANE persists even without long-range order, provided that local symmetry breaking retains statistical significance. Additionally, the pseudogap observed in transport [47] above $T_{CDW}$ may sustain the ANE without long-range coherence. Notably, near $T_{CDW}$, $S_{xz}^A$ undergoes a sudden increase, reaching a maximum value of ~ 4 μV/K. This value is comparable to those of topological magnets[8, 48, 49]. This enhancement of $S^A_{xz}$ completely coincides with the coexistence region of the S-CDW and L-CDW in Figure 4i. This observation suggests that the interplay of these competing CDW orders is also crucial in governing the resulting behavior of ANE.

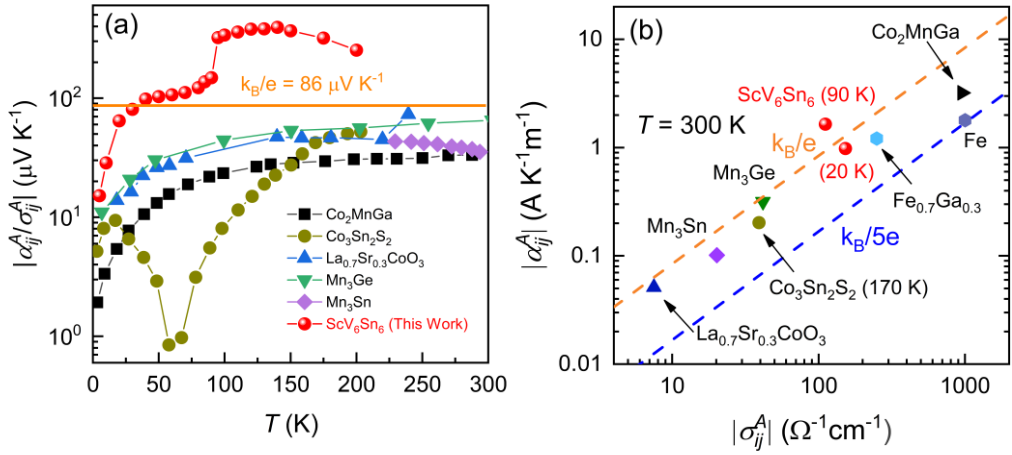

Figure 4 (a) The $|\alpha_{ij}^A / \sigma_{ij}^A|$ ratio as a function of temperature in different magnets and ScV$_6$Sn$_6$. (b) Comparing $|\alpha_{ij}^A / \sigma_{ij}^A|$ in nonmagnetic metal ScV$_6$Sn$_6$ with those in other typical magnetic materials. They lie between $k_B/5e$ (blue line) and $k_B/e$ (red line).

We can now turn our attention to the $|\alpha_{ij}^A / \sigma_{ij}^A|$ ratio in ScV$_6$Sn$_6$ and compare it with available data from typical magnetic system[50], as shown in Figure 4a. The ratio relates the Peltier coefficient, which denotes the transport of entropy, to the Hall coefficient, which is indicative of charge transport. Among these systems, the non-magnetic ScV$_6$Sn$_6$ stands out due to its obvious jump at $T_{CDW}$ and the higher $|\alpha_{ij}^A / \sigma_{ij}^A|$ values. Previous study on CoMnGa$_2$ [50] has argued that this ratio of $|\alpha_{ij}^A / \sigma_{ij}^A|$ can be limited by a saturated value related to the constant $k_B/e = 86$ μV/K. This value indicates that the anomalous transverse transport effects, including AHE and ANE, arise from the contribution of Berry curvature. In contrast, the $|\alpha_{ij}^A / \sigma_{ij}^A|$ ratio in ScV$_6$Sn$_6$ is notably found to surpass the bound value at a temperature range from 40 K to 200 K. This indicates that additional mechanisms like multiple bands and charge order fluctuations can be involved to simultaneously contribute to the Hall and Nernst responses. Similar cases have been observed in other topological magnets, such as UCo$_{0.8}$Ru$_{0.2}$Al[51]. Figure 4b further demonstrates the intrinsic nature of $\alpha_{ij}^A$ in ScV$_6$Sn$_6$ by comparing it with that of typical magnetic materials. Despite a 100-fold variation in the $\sigma_{ij}^A$ across different topological materials, the room temperature $|\alpha_{ij}^A / \sigma_{ij}^A|$ ratio consistently falls within the universal range of $k_B/5e$ and $k_B/e$ -the natural units of the ratio of these correlated quantities. Notably, ScV$_6$Sn$_6$ exhibits slight deviations from this trend, at 90 K, its $\sigma_{ij}^A$ exceeds slightly the $k_B/e$ boundary, while at 20 K, it aligns with the expected window. This feature underscores the dominant role of Berry curvature in governing the anomalous transverse coefficients. Similar trends have been observed in other topological materials, including Co$_3$Sn$_2$S$_2$[8], CoMnGa$_2$[50], and Mn$_3$Sn[52]. This larger $\alpha_{ij}^A$ in ScV$_6$Sn$_6$ near $T_{CDW}$ may imply the charge fluctuations and competing CDW states may enhance the anomalous thermoelectric effect. Our findings highlight the potential of ScV$_6$Sn$_6$ for applications in thermoelectric devices, especially given its strong performance amid competing CDW states. These results encourage further exploration into the underlying mechanisms that govern the observed thermoelectric response and its relationship to CDW orderings.

## Conclusion

In summary, we observed unconventional anomalous thermoelectric properties associated with charge order, topological bands, and hidden magnetism in the kagome metal ScV$_6$Sn$_6$. Below $T_{CDW}$, the thermal conductivity and thermopower respond sensitively to applied magnetic fields, suggesting a potential presence of hidden magnetism. Importantly, we report the first observation of a step-like Nernst signal in ScV$_6$Sn$_6$, manifesting an anomalous Nernst component that peaks a value of ~ 4 µV/K around $T_{CDW}$. These observations indicate that ANE likely stems from hidden magnetism as one plausible origin of TRSB in the CDW state. However, we cannot exclude alternative origins involving chiral charge order or orbital currents, both of which could similarly generate the observed transport signatures. Further studies combining local probe techniques and theoretical modeling will be essential to unambiguously clarify the intrinsic mechanism. The observed magneto-thermopower oscillations further support the existence of non-trivial topological bands. Importantly, the anomalous Nernst signals persist at higher temperatures above $T_{CDW}$, suggesting possible contributions from the competition with short-range CDW orders. Overall, our study highlights the unconventional anomalous thermoelectric properties associated with time-reversal symmetry breaking in the non-magnetic Kagome compound ScV$_6$Sn$_6$, emphasizing its potential for exploring novel thermoelectric phenomena in topological materials.

**Acknowledgments**
This work was supported by the Hangzhou Joint Fund of the Zhejiang Provincial Natural Science Foundation of China (under Grants No. LHZSZ24A040001) and the Nature Science funding (NSF) of China (under Grants No. U1932155, 12274109, 12304175). The work at Zhejiang University acknowledges support from the National Key R&D Program of China (No. 2022YFA1402200, 2024YFA1409202), the Key R&D Program of Zhejiang Province, China (2021C01002), the National Natural Science Foundation of China (No. 12350710785, 12274363, 12274364), and the Fundamental Research Funds for the Central Universities (Grant No. 2021QNA3003).

# The authors contributed equally to this work.
Corresponding authors: yusong_phys@zju.edu.cn; yklee@hznu.edu.cn

**Competing interests:** The authors declare that they have no competing interests.

# References

[1] Y. Wang, H. Wu, G.T. McCandless, J.Y. Chan, M.N. Ali, Nature Reviews Physics. **5**, 635-658 (2023).
[2] J.-X. Yin, B. Lian, M.Z. Hasan, Nature. **612**, 647-657 (2022).
[3] T. Neupert, M.M. Denner, J.-X. Yin, R. Thomale, M.Z. Hasan, Nature Physics. **18**, 137-143 (2021).
[4] B.R. Ortiz, L.C. Gomes, J.R. Morey, M. Winiarski, M. Bordelon, J.S. Mangum, I.W.H. Oswald, J.A. Rodriguez-Rivera, J.R. Neilson, S.D. Wilson, et al., Physical Review Materials. **3**, 094407 (2019).
[5] X. Teng, L. Chen, F. Ye, E. Rosenberg, Z. Liu, J.-X. Yin, Y.-X. Jiang, J.S. Oh, M.Z. Hasan, K.J. Neubauer, et al., Nature. **609**, 490-495 (2022).
[6] H.W.S. Arachchige, W.R. Meier, M. Marshall, T. Matsuoka, R. Xue, M.A. McGuire, R.P. Hermann, H. Cao, D. Mandrus, Physical Review Letters. **129**, 216402 (2022).
[7] Y.-X. Jiang, J.-X. Yin, M.M. Denner, N. Shumiya, B.R. Ortiz, G. Xu, Z. Guguchia, J. He, M.S. Hossain, X. Liu, et al., Nature Materials. **20**, 1353-1357 (2021).
[8] H. Yang, W. You, J. Wang, J. Huang, C. Xi, X. Xu, C. Cao, M. Tian, Z.-A. Xu, J. Dai, et al., Physical Review Materials. **4**, 024202 (2020).
[9] X. Zhou, H. Liu, W. Wu, K. Jiang, Y. Shi, Z. Li, Y. Sui, J. Hu, J. Luo, Physical Review B. **105**, 205104 (2022).
[10] S.-Y. Yang, Y. Wang, B.R. Ortiz, D. Liu, J. Gayles, E. Derunova, R. Gonzalez-Hernandez, L. Šmejkal, Y. Chen, S.S.P. Parkin, et al., Science Advances. **6**, eabb6003 (2020).
[11] Q. Yin, Z. Tu, C. Gong, S. Tian, H. Lei, Chinese Physics Letters. **38**, 127401 (2021).
[12] J. Bernhard, B. Lebech, O. Beckman, Journal of Physics F: Metal Physics. **14**, 2379 (1984).
[13] C. Shi, Y. Liu, B.B. Maity, Q. Wang, S.R. Kotla, S. Ramakrishnan, C. Eisele, H. Agarwal, L. Noohinejad, Q. Tao, et al., Science China Physics, Mechanics & Astronomy. **67**, 117012 (2024).
[14] J.J. Ma, C.F. Shi, Y.T. Cao, Y.W. Zhang, Y.Z. Li, J.X. Liao, J.L. Wang, W.H. Jiao, H.J. Guo, C.C. Xu, et al., Science China Physics, Mechanics & Astronomy. **68**, 237412 (2025).
[15] X. Wu, X. Mi, L. Zhang, C.-W. Wang, N. Maraytta, X. Zhou, M. He, M. Merz, Y. Chai, A. Wang, Physical Review Letters. **132**, 256501 (2024).
[16] Z. Chen, X. Wu, S. Zhou, J. Zhang, R. Yin, Y. Li, M. Li, J. Gong, M. He, Y. Chai, et al., Nature Communications. **15**, 6262 (2024).
[17] H. Tan, B. Yan, Physical Review Letters. **130**, 266402 (2023).
[18] S. Lee, C. Won, J. Kim, J. Yoo, S. Park, J. Denlinger, C. Jozwiak, A. Bostwick, E. Rotenberg, R. Comin, et al., npj Quantum Materials. **9**, 15 (2024).
[19] A. Korshunov, H. Hu, D. Subires, Y. Jiang, D. Călugăru, X. Feng, A. Rajapitamahuni, C. Yi, S. Roychowdhury, M.G. Vergniory, et al., Nature Communications. **14**, 6646 (2023).
[20] R. Guehne, J. Noky, C. Yi, C. Shekhar, M.G. Vergniory, M. Baenitz, C. Felser, Nature

Communications. **15**, 8213 (2024).

[21] S. Cao, C. Xu, H. Fukui, T. Manjo, Y. Dong, M. Shi, Y. Liu, C. Cao, Y. Song, Nature Communications. **14**, 7671 (2023).

[22] J.-X. Yin, W. Ma, T.A. Cochran, X. Xu, S.S. Zhang, H.-J. Tien, N. Shumiya, G. Cheng, K. Jiang, B. Lian, et al., Nature. **583**, 533-536 (2020).

[23] T. Hu, H. Pi, S. Xu, L. Yue, Q. Wu, Q. Liu, S. Zhang, R. Li, X. Zhou, J. Yuan, et al., Physical Review B. **107**, 165119 (2023).

[24] S. Cheng, Z. Ren, H. Li, J.S. Oh, H. Tan, G. Pokharel, J.M. DeStefano, E. Rosenberg, Y. Guo, Y. Zhang, et al., npj Quantum Materials. **9**, 14 (2024).

[25] Z. Guguchia, D.J. Gawryluk, S. Shin, Z. Hao, C. Mielke Iii, D. Das, I. Plokhikh, L. Liborio, J.K. Shenton, Y. Hu, et al., Nature Communications. **14**, 7796 (2023).

[26] C. Yi, X. Feng, N. Mao, P. Yanda, S. Roychowdhury, Y. Zhang, C. Felser, C. Shekhar, Physical Review B. **109**, 035124 (2024).

[27] S. Mozaffari, W.R. Meier, R.P. Madhogaria, N. Peshcherenko, S.-H. Kang, J.W. Villanova, H.W.S. Arachchige, G. Zheng, Y. Zhu, K.-W. Chen, et al., Physical Review B. **110**, 035135 (2024).

[28] C. Yi, X. Feng, N. Kumar, C. Felser, C. Shekhar, New Journal of Physics. **26**, 052001 (2024).

[29] G. Pokharel, B.R. Ortiz, L. Kautzsch, S.J. Gomez Alvarado, K. Mallayya, G. Wu, E.-A. Kim, J.P.C. Ruff, S. Sarker, S.D. Wilson, Physical Review Materials. **7**, 104201 (2023).

[30] A. Subedi, Physical Review Materials. **8**, 014006 (2024).

[31] Y. Hu, J. Ma, Y. Li, Y. Jiang, D.J. Gawryluk, T. Hu, J. Teyssier, V. Multian, Z. Yin, S. Xu, et al., Nature Communications. **15**, 1658 (2024).

[32] Y. Li, L. Li, J. Wang, T. Wang, X. Xu, C. Xi, C. Cao, J. Dai, Physical Review B. **94**, 121115 (2016).

[33] G. Zheng, Y. Zhu, S. Mozaffari, N. Mao, K.-W. Chen, K. Jenkins, D. Zhang, A. Chan, H.W.S. Arachchige, R.P. Madhogaria, et al., Journal of Physics: Condensed Matter. **36**, 215501 (2024).

[34] E. Rosenberg, J. DeStefano, Y. Lee, C. Hu, Y. Shi, D. Graf, S.M. Benjamin, L. Ke, J.-H. Chu, arXiv preprint arXiv: 2401.14699 (2024).

[35] K. Yang, W. Xia, X. Mi, L. Zhang, Y. Gan, A. Wang, Y. Chai, X. Zhou, X. Yang, Y. Guo, et al., Physical Review B. **107**, 184506 (2023).

[36] T. Liang, J. Lin, Q. Gibson, T. Gao, M. Hirschberger, M. Liu, R.J. Cava, Physical Review Letters. **118**, 136601 (2017).

[37] X. Chen, X. Liu, W. Xia, X. Mi, L. Zhong, K. Yang, L. Zhang, Y. Gan, Y. Liu, G. Wang, et al., Physical Review B. **107**, 174510 (2023).

[38] Y. Shu, X. Mi, Y. Wei, S. Tao, A. Wang, Y. Chai, D. Ma, X. Yang, M. He, Physical Review B. **111**, 155103 (2025).

[39] Y. Gan, W. Xia, L. Zhang, K. Yang, X. Mi, A. Wang, Y. Chai, Y. Guo, X. Zhou, M. He, Physical Review B. **104**, L180508 (2021).

[40] D. Chen, B. He, M. Yao, Y. Pan, H. Lin, W. Schnelle, Y. Sun, J. Gooth, L. Taillefer, C. Felser, Physical Review B. **105**, L201109 (2022).

[41] H. Yang, Q. Wang, J. Huang, Z. Wang, K. Xia, C. Cao, M. Tian, Z. Xu, J. Dai, Y. Li, Science China Physics, Mechanics & Astronomy. **65**, 117411 (2022).

[42] R. Ramos, M.H. Aguirre, A. Anadón, J. Blasco, I. Lucas, K. Uchida, P.A. Algarabel, L. Morellón, E. Saitoh, M.R. Ibarra, Physical Review B. **90**, 054422 (2014).

[43] A. Sakai, Y.P. Mizuta, A.A. Nugroho, R. Sihombing, T. Koretsune, M.-T. Suzuki, N. Takemori, R. Ishii, D. Nishio-Hamane, R. Arita, et al., Nature Physics. **14**, 1119-1124 (2018).

[44] S. Roychowdhury, A.M. Ochs, S.N. Guin, K. Samanta, J. Noky, C. Shekhar, M.G. Vergniory, J.E. Goldberger, C. Felser, Advanced Materials. **34**, 2201350 (2022).

[45] S. Liu, C. Wang, S. Yao, Y. Jia, Z. Zhang, J.-H. Cho, Physical Review B. **109**, L121103 (2024).

[46] M. Tuniz, A. Consiglio, D. Puntel, C. Bigi, S. Enzner, G. Pokharel, P. Orgiani, W. Bronsch, F. Parmigiani, V. Polewczyk, et al., Communications Materials. **4**, 103 (2023).

[47] J.M. DeStefano, E. Rosenberg, O. Peek, Y. Lee, Z. Liu, Q. Jiang, L. Ke, J.-H. Chu, npj Quantum Materials. **8**, 65 (2023).

[48] Y. Pan, F.R. Fan, X. Hong, B. He, C. Le, W. Schnelle, Y. He, K. Imasato, H. Borrmann, C. Hess, et al., Advanced Materials. **33**, 2003168 (2020).

[49] M. Ikhlas, T. Tomita, T. Koretsune, M.-T. Suzuki, D. Nishio-Hamane, R. Arita, Y. Otani, S. Nakatsuji, Nature Physics. **13**, 1085-1090 (2017).

[50] L.C. Xu, X.K. Li, L.C. Ding, T.S. Chen, A. Sakai, B. Fauqu, S. Nakatsuji, Z.W. Zhu, K. Behnia, Physical Review B. **101**, 180404 (2020).

[51] T. Asaba, V. Ivanov, S.M. Thomas, S.Y. Savrasov, J.D. Thompson, E.D. Bauer, F. Ronning, Science Advances. **7**, eabf1467 (2021).

[52] X. Li, L. Xu, L. Ding, J. Wang, M. Shen, X. Lu, Z. Zhu, K. Behnia, Physical Review Letters. **119**, 056601 (2017).